\documentstyle[12pt]{article}

\renewcommand{\arraystretch}{1.5}

\newcommand{\beq}{\begin{equation}}
\newcommand{\eeq}{\end{equation}}
\newcommand{\bea}{\begin{eqnarray}}
\newcommand{\eea}{\end{eqnarray}}
\newcommand{\bi}{\bibitem}
\newcommand{\p}{\partial}

\renewcommand{\theequation}{\arabic{section}.\arabic{equation}}
\renewcommand{\thesection}{\arabic{section}.}

\begin{document}

\begin{flushright}
                  {\tt DIAS-STP 92}--{\tt 24} \\ 
                  {\tt gr-qc/9211011}
\end{flushright}
\vspace{0.5cm}

\begin{center}
              {\LARGE Weak-Field Gravity of Circular Cosmic Strings}
\end{center}

\begin{center}{\Large
		     ${}^{\ast}$Shane J. Hughes,
                     ${}^{\dag}$Des J. Mc Manus}
\end{center}

\begin{center}{\sl
                 School of Theoretical Physics,
                        Dublin Institute for Advanced Studies,\\
                        10 Burlington Road, Dublin 4, Ireland.
              }
\end{center}

 \vskip 5 truemm

{\Large \centerline{${}^{\ddag}$Michel A. Vandyck}}

\begin{center}{\sl
                  Physics Department, University College
                        Cork, Ireland, \\ and \\
                  Physics Department, Cork Regional Technical
                  College, Ireland.
               }
\end{center}

\vspace{.5cm}

%

\begin{center}{\large {\bf Abstract}} \end{center}

\noindent
   A weak-field solution of Einstein's equations is
   constructed. It is generated by
   a circular cosmic string externally supported against collapse.
   The solution exhibits a conical singularity, and the corresponding
   deficit angle is the same as for a straight string of the same
   linear energy density. This confirms the deficit-angle assumption
   made in the Frolov-Israel-Unruh derivation of the metric describing
   a string loop at a moment of time symmetry.

 \noindent PACS numbers:  04.20-q, 98.90+s

\vspace{1cm}

\begin{center}{\it
                To appear in Physical Review D}
\end{center}

\pagebreak

                          \section{Introduction}                              %

In a gauge theory with gauge group $G$ spontaneously broken via the
Higgs mechanism to a subgroup $H$, the Higgs field must lie on the
manifold of degenerate vacua $G/H$ on the circle at spatial
infinity. Thus if $\pi_1(G/H)$ is non trivial, string-like
configurations with finite energy per unit length
will exist for which the Higgs field at spatial infinity winds
non-trivially around $G/H$. Such so-called cosmic strings have received
much attention in recent years, especially as providing possible
mechanisms for the formation of large-scale
structure in the universe \cite{1}. Accretion wakes have  been proposed
to form behind infinitely long relativistic strings,   which
may help explain the observed large-scale filaments.
Moreover, Vilenkin \cite{2}  has suggested that long-lived closed loops of
string might act as seeds for galaxy and cluster formation.

The derivation of the gravitational field outside a
cosmic string, in the weak-field approximation,
was carried out by Vilenkin \cite{3}, who investigated the
case of an infinitely thin straight string. The main features of this
solution are that the spacetime exterior to the string
is flat, and that all along the string, the spacetime has a conical
singularity with deficit angle $\delta\phi = 8\pi G \mu$, where $\mu$ is
the linear energy density. (The solution to the exact Einstein equations
have since been derived \cite{exact}.)

Closed string loops have been studied by Frolov, Israel, and
Unruh (FIU) \cite{FIU}. Using the initial-value formulation \cite{ADM},
they  constructed a family of momentarily stationary
circular loops, which are considered as thin
loops of string either at the time of formation or at the
turning point between expansion and collapse.
One of the characteristics of the FIU method, which has the
merit of taking into account the
exact non-linear field equations, is that it imposes {\it a priori} that
all points on the circular string be conical singularities of
spacetime with deficit angle $\delta\phi = 8\pi G \mu$. This is a
reasonable assumption because of the observation \cite{FIU}
that an infinitely thin circular string, when viewed from
arbitrarily close to the core, is indistinguishable from
an infinitely thin straight string, and that
all the points on the circular string may thus be assumed to be  conical
singularities with the same angular deficit
as for a straight string of equal linear energy density.

The question we ask is the following:
instead of making the above assumption,
can one deduce from the field equations, at least
in the weak field limit, the fact that all the points on the circular string
are
conical singularities? As a first step in this direction one may
construct, without any special hypothesis on conical singularities,
a solution produced by a circular source
generated by a stress-energy tensor
obtained by adapting the method used by Vilenkin \cite{3} for a
straight string.  This solution could then be examined for possible
conical singularities and the corresponding angular
deficits, if any, could be calculated.
We carry out such an investigation in this paper.

An important difference between our analysis and that of FIU is
that we seek a solution which is stationary, whereas FIU
consider a circular string loop which is momentarily at rest.
Conceivably, a non-stationary solution at a moment
of time symmetry may differ from a stationary solution by the
presence of gravitational waves, but FIU considered precisely
a class of particular solutions devoid of such free gravitational
radiation.  Therefore, the comparison between the two cases is
physically reasonable. Of course an isolated stationary circular
string is
unphysical, since it would tend to collapse. To overcome this
problem,
radial stresses, the values of which are determined by
stress-energy conservation conditions, are introduced to support
the string. It will be seen that these stresses do not contribute to
the value of the angular deficit.

In section two, we establish the appropriate form of the stress-energy tensor
for this problem. We derive and solve the weak-field Einstein equations
in sections three and four. In section five, we demonstrate
that the solution does indeed exhibit conical singularities with the
same angular deficit as for a straight string with the same linear
energy density, thus fully supporting and agreeing with the FIU
hypothesis.

In this paper, we use units in which $\hbar = c = 1$, take the
metric to have signature $(-,+,+,+)$ and adopt the geometrical
conventions of \cite{SYNGE}.

           \section{The Stress-Energy Tensor and the Metric}                  %

In this section, we establish the stress-energy tensor ${T_\mu}^\nu$
for a loop of cosmic string arising from a spontaneously broken
gauge theory. In the case of an infinitely long straight string, all
components of the stress-energy tensor are localised on the core of
the string, that is the region of spacetime where the Higgs field
unwinds and thus does not lie on the manifold of degenerate vacua.
The radius of this region is of the order of the Compton wavelength
of the Higgs field \cite{PRESKILL}, generically a microscopic distance.
Thus, when considering the gravitational effects of the string, it is
physically reasonable to make the thin-string approximation where the
stress-energy tensor is localised on an infinitely thin line.

A circular source produces an axially symmetric gravitational
field, and the most general stationary metric of this type
may be cast in the form \cite{SYNGE}
\beq
    ds^2 \;=\; -e^{2\nu}\,dt^2 \;+\;  e^{2\zeta - 2\nu}\, r^2\,
    d\phi^2 \;+\; e^{2\eta - 2\nu} \,(dr^2\,+\, dz^2)\;\;\;,
    \label{2_1}
\eeq
where the three functions $\nu, \zeta$, and $\eta$ depend only on $r$ and $z$,
and $x^\alpha \,\equiv\, (t, \phi, r, z)$ denotes cylindrical  coordinates.

The form of the Einstein tensor for this metric, together with
the Einstein field equations $G_{\mu}{}^{\nu} = -\kappa T_{\mu}{}^{\nu}$
($\kappa = 8\pi G$) implies that the most general stress-energy
tensor compatible with (\ref{2_1}) is
\renewcommand{\arraystretch}{1}
\beq
   {T_\mu}^\nu \,=\, \left(
    \begin{array}{cccc}
           \alpha & 0 & 0 & 0 \\
           0 & \gamma & 0 & 0 \\
           0 & 0 & \Delta & \epsilon \\
           0 & 0 & \epsilon & \beta
    \end{array} \right) \;\;\;, \label{2_2}
\eeq
\renewcommand{\arraystretch}{1.5} 
where $\alpha, \beta, \gamma, \Delta$, and $\epsilon$
are solely functions of $r$ and $z$.

In passing to the limit of an infinitely thin
 loop of string with radius $a$, lying in
the $x$-$y$ plane and centred at the origin,
we obtain (with the exception of ${T_r}^r$, dealt with below)
the stress-energy tensor as
\beq
    {T_\mu}^\nu = {\bar{T}}_\mu{}^\nu \,
    \delta(r-a) \, \delta(z)
    \;\;\;,
\eeq
where ${\bar{T}_\mu}{}^\nu$ is the cross-sectional
integral of ${T_\mu}^\nu$.
Therefore this limit yields, in an obvious notation:
\bea
    \alpha \;&=&\; \bar{\alpha} \, \delta (r-a) \, \delta (z)  \\
    \beta \;&=&\; \bar{\beta}\, \delta (r-a) \, \delta(z)     \\
    \gamma \;&=&\; \bar{\gamma} \, \delta (r-a) \, \delta (z)
    \label{2_4} \\
    \epsilon \;&=&\; \bar{\epsilon} \, \delta (r-a)\, \delta (z) \;\;\;.
\eea
However, such a localisation does not occur
for the radial stress ${T_r}^r \equiv \Delta$ required to support
the ring. Given that the ring is {\it externally} supported, this radial
stress is not confined to the core. (In what follows, we choose to
support the ring from infinity.)

Following Vilenkin, we can show that
$\bar{\beta}=\bar{\epsilon}=0$.
Indeed, taking an arbitrary string cross-section
${\cal M} = \{ (\phi, r, z) : \phi =$ const$\}$ and using
stress-energy conservation $({T_\mu}^{\nu}{}_{\mid \nu} = 0)$, we have
\beq
    0 \,=\, \int_{\cal M} {T_\mu}^\nu \!_{\mid\nu} \, x^\lambda \, dr dz
\;\;\;.
\eeq
In the weak-field limit, the components of the stress-energy
tensor (\ref{2_2})
are taken to be of the same order of magnitude as the metric
functions $\nu, \zeta,$ and $\eta$, which are assumed to be small
compared to unity. Thus to lowest order, the only Christoffel
symbols for the metric (\ref{2_1}) that contibute to the covariant
derivative of the stress-energy tensor are
${\Gamma^r}_{\phi\phi} =-r$ and
${\Gamma^\phi}_{r\phi}={\Gamma^\phi}_{\phi r}= 1/r$.
Taking $\mu = z$ and noting that ${T_z}^t = {T_z}^\phi = 0$, we
may integrate by parts to get
\beq
    0 \,=\,
         \bar{T}_z{}^r \; \delta_r^\lambda \;\;
         + \;\; \bar{T}_z{}^z \; \delta_z^\lambda
          \;\;\;, \label{2_2a}
\eeq
(no sum over $r$ and $z$).
There is no boundary contribution since ${T_z}^{\lambda}$ is localised
on the string core. We emphasise that, in contrast with the case of
the straight string, a similar argument would fail to show that
$\bar{T}_r{}^r=0$ because ${T_r}^r \equiv \Delta$ is non-zero
outside the core. We obtain
the desired result ($\bar{\epsilon} = \bar{\beta} =0$)
by taking $\lambda = r, z$ in (\ref{2_2a}).
The last undetermined function, $\Delta$, in the stress-energy tensor
can be found in terms of $\gamma$  from  the stress-energy
conservation equations, and will be calculated in the next
section.

At this stage, we have deduced that the stress-energy tensor for a
thin circular matter source supported by external radial stresses
is given by (\ref{2_2}) with $\beta = \epsilon = 0$, and $\alpha$,
$\Delta$ and $\gamma$ as above. We rewrite $ \bar{\alpha} = -\mu$ and
$ \bar{\gamma} = k$,
where $\mu$ and $k$ are the linear energy density
and the azimuthal stress per unit length, respectively.
Henceforth, we make the thin-string approximation and use this
form of the stress-energy tensor.

To specialise to the case of a cosmic string in a spontaneously
broken gauge theory, we would apply the equation of state \cite{3} for string
matter $k= -\mu$. (This equation of state is dictated by
Lorentz invariance in the straight-string case; for a circular string, the
azimuthal stress ${T_\phi}^\phi$ plays the same role as that of the
longitudinal stress ${T_z}^z$ for a straight string.)
We will, however, keep $\mu$ and $k$ as two independent parameters.
This has the advantage that we will then be able to compare
and contrast ordinary matter, given by $k$ positive and small
compared to $\mu$, with string matter, given by $k = -\mu$.

\setcounter{equation}{0} 

       \section{The Field Equations and Stress-Energy Conservation}         %

In the weak-field approximation, the non-zero
components of the Einstein tensor\cite{C&F} for the metric (\ref{2_1})
are, upon retaining
only first-order terms in $\nu,\zeta$ and $\eta$:
\bea
    {G_t}^t &=&2\nabla^2\,\nu \,-\, \nabla^2\, \zeta \,-\,
    \frac{1}{r} \,\zeta_r \,-\, {\widetilde{\nabla}}^2\, \eta \\
    {G_\phi}^\phi &=&-{\widetilde{\nabla}}^2\, \eta \\
    {G_r}^r &=&-\zeta_{zz} \,-\, \frac{1}{r} \,\eta_r \\
    {G_z}^z &=&-\zeta_{rr} \,+\, \frac{1}{r} \,(\eta_r \,-\,2\,\zeta_r) \\
    {G_r}^z &=&\zeta_{zr} \,+\, \frac{1}{r} \,(\zeta_z \,-\, \eta_z) \;\;\;,
\eea
where ${\widetilde{\nabla}}^2\, \equiv \p^2_r + \p^2_z$, and
$\nabla^2\, \equiv \p^2_r + \frac{1}{r} \p_r + \p^2_z$ is the
flat-space Laplacian operator.
(We stress that by \lq\lq weak-field approximation,\rq\rq\   we mean
retaining only first-order terms in $\nu, \eta$ and $\zeta$.)
After a short calculation, the field
equations reduce to
\bea
    \nabla^2 \nu &=& 4\pi G \,(-\alpha \,+\,\gamma\,+\, \Delta) \label{3_1} \\
    {\widetilde{\nabla}}^2\, \eta &=& 8\pi G \,\gamma  \label{3_2} \\
    \nabla^2\, \zeta \,+\, \frac{1}{r} \,\zeta_r &=& 8\pi G \,
                                                   \Delta \label{3_3} \\
    \eta_r &=& r\, \zeta_{rr} \,+\, 2 \zeta_r \label{3_4} \\
    \eta_z &=& r\, \zeta_{zr} \,+\,  \zeta_z \label{3_5}   \;\;\;.
\eea
We integrate (\ref{3_4}) and (\ref{3_5})  to get
\bea
    \eta &=& \frac{\partial}{\partial r} (r\zeta) \,+\, \eta_0 \label{3_6}
    \;\;\;,
\eea
where $\eta_0$ is an arbitrary constant to be determined later.

The conservation of energy and momentum
[or equivalently the compatibility conditions
for (\ref{3_2})--(\ref{3_5})] places constraints on
the allowable components of ${T_\mu}^\nu\;$,
which in the weak-field limit read
\beq
    \gamma \,=\, \frac{\partial}{\partial r}(r\Delta)   \label{3_7} \;\;\;.
\eeq
The solution of (\ref{3_7}) for $\Delta$, with the azimuthal stress $\gamma$
given by (\ref{2_4}) and corresponding to supporting the string  from
infinity,  is
\beq
    \Delta(r, z) \;=\;\frac{k}{r} \; \Theta(r-a) \;
    \delta(z) \;\;\;, 
\eeq
in which $\Theta$  denotes the Heaviside
step function.  Furthermore, (\ref{3_6})
and (\ref{3_7}) imply that (\ref{3_2}) and (\ref{3_3})
are equivalent. We may therefore discard (\ref{3_3}), retain (\ref{3_2})
to find $\eta$, and use (\ref{3_6}) to express $\zeta$
in terms of $\eta$.

Combining all the results, the field equations are equivalent to
solving the following set of equations for
$\nu$, $\eta$ and $\zeta$:
\bea
    \nabla^2\,\nu \;&=&\; 4\pi G \, [(\mu + k)\, \delta(r-a) \,+\,
                       \frac{k}{r} \, \Theta(r-a)]\,\delta(z) \label{3_9} \\
    {\widetilde{\nabla}}^2\, \eta \;&=&\; 8\pi G\, k \,
    \delta(r-a) \, \delta(z)
    \label{3_10} \\
    \frac{\partial}{\partial r}(r\zeta) \;&=&\; \eta \,-\, \eta_0
    \label{3_11}    \;\;\;.
\eea
These equations involve the dimensionless quantities $G\mu$ and $Gk$.
For a cosmic string, $k = -\mu$, whereas for ordinary matter, $k <<
\mu$. The largest value of $\mu$ in a physically relevant theory
occurs for GUT strings, for which $G\mu \approx 10^{-6}$ \cite{3},
validating the
weak-field approximation as an expansion of $\nu, \zeta$, and $\eta$
in powers of $G\mu$.

\setcounter{equation}{0}  

                   \section{Solution of the Field Equations}
%

 We note that (\ref{3_9}) is Poisson's equation for the Newtonian
potential $\nu$. The solution is most easily found \cite{BATEMAN}
in toroidal coordinates $(\phi ,\sigma, \psi)$. The latter are related to
cylindrical coordinates $(\phi,r,z)$ by
\beq
    \begin{array}{ccccc}
           z \,=\, aN^{-2}\, \sin \psi & \;\;\;\;\;\;& r\,=\,
           aN^{-2}\, {\rm sh} \sigma & \;\;\;\;\;\; & ( 0 \leq \sigma \leq
           \infty \;,\, \mid\! \psi \! \mid \leq \pi) \;\;\;,
    \end{array}  \label{9957}
\eeq
where $N^2 \equiv N^2(\sigma, \psi) \equiv  {\rm ch} \sigma - \cos \psi$.
The surfaces
$\sigma = \sigma_0 \equiv$ const. are tori whose generating circles
have radii $a\,$cosech$\,\sigma_0$ and
$a \coth \sigma_0$. In particular, the ring $r=a, \, z=0$ is now
given by $\sigma = \infty$. (See Fig. 1 in  \cite{FIU}.)
In toroidal coordinates, the metric (\ref{2_1}) reads
\beq
    ds^2 \,=\, -e^{2\nu}\, dt^2 \,+\, a^2\, N^{-4}\, {\rm sh}^2\sigma
    \; e^{2\zeta - 2 \nu}\, d\phi^2 \,+\, a^2\, N^{-4}\, e^{2\eta - 2 \nu}
    \, (d\sigma^2 \,+\, d\psi^2) \label{20}\;\;\;,
\eeq
where $\nu, \zeta$, and $\eta$ depend on $\sigma$ and $\psi$.

The potential $\nu$ can be split into two parts,  $\nu =\nu_1 +\nu_2$,
where $\nu_1$ and $\nu_2$ satisfy separate Poisson equations:
\bea
    \nabla^2 \,\nu_1 &=& 4\pi G\,(\mu + k) \, \delta(r-a)\,\delta(z)
    \label{4_1} \\
    \nabla^2 \, \nu_2 &=& 4\pi G\, \frac{k}{r} \,\Theta(r-a)\,\delta(z)
    \label {4_2} \;\;\;.
\eea
The function $\nu_1$ is the gravitational potential produced by
a circular ring of matter. This problem is known \cite{BATEMAN},
and the corresponding solution is
\beq
\nu_1 \;=\; -2^{3/2} \, G(\mu + k) \,N(\sigma, \psi)\,
K({\rm th} (\frac{1}{2}\sigma)) \, / {\rm ch}(\frac{1}{2}\sigma)
\;\;\;,\label{4_3}
\eeq
where  $K$ denotes the Complete Elliptic Integral of the First Kind.

The formal solution of Poisson's equation  with source $\rho$
is given by
\beq
    \nu_2 ({\bf\vec{r}}) \,=\,-G
\int \frac{\rho({\bf {\vec{r}}_0}) \,dV_0}{\mid\! {\bf \vec{r}} -
    {\bf \vec{r}_0} \!\mid}       \;\;\;.
    \label{23}
\eeq
Moreover, the Green's function in toroidal
coordinates has the expansion \cite{M&F}
\[
          \frac{\pi a}{\mid\! {\bf \vec{r}} - {\bf\vec{r}_0} \!\mid}  \,=\,
          N(\sigma,\psi) \, N(\sigma_0, \psi_0)
           \, \sum^{\infty}_{m,n = 0} C_{mn} \,
          \cos m(\phi - \phi_0) \, \cos n(\psi - \psi_0)
\]
\beq
    \;\;\;\;\;\;\;\;\;\;\;\;\;\;\;\;\;\;\;\;\;\;\;\;\;\; \times
          P^m_{n-\frac{1}{2}}({\rm ch} \sigma_{<}) \,
          Q^m_{n-\frac{1}{2}}({\rm ch} \sigma_{>}) \;\;\;,
    \label{24}
\eeq
where $\sigma_{>} \equiv \max\{\sigma, \sigma_0\}$, $\sigma_{<} \equiv
\min\{\sigma, \sigma_0\}$, and the functions $P^m_{n-\frac{1}{2}}$
and $Q^m_{n - \frac{1}{2}}$ are toroidal Legendre functions \cite{A&S}.
The numerical coefficients are
\beq
    C_{mn} \,=\, (-1)^m \,
    \frac{\Gamma(n-m+\frac{1}{2})}{\Gamma(n+m+\frac{1}{2})} \,
    \varepsilon_n \, \varepsilon_m \label{25} \;\;\;,
\eeq
where $\varepsilon_n \equiv 2 - \delta^0_n$.

In toroidal coordinates, we have
\beq
    \Theta(r-a)\,\delta(z) \,=\,  a^{-1} \,N^2(\sigma, \psi) \, \delta(\psi)
    \;\;\;. \label{4_t}
\eeq
Inserting (\ref{4_t}), together with (\ref{24}) and  (\ref{25}),
into (\ref{23}), we find the following expression for $\nu_2$ in
toroidal coordinates:
\beq
    \nu_2 \,=\,-2 G k \,N(\sigma,\psi)
     \, \sum^{\infty}_{n=0} \left[ F_n(\sigma) \,
              P_{n-\frac{1}{2}}({\rm ch} \sigma) \,+\, G_n(\sigma) \,
              Q_{n-\frac{1}{2}}({\rm ch} \sigma) \right]
\varepsilon_n\,\cos n\psi
              \label{27}\;\;\;,
\eeq
where
\bea
    F_n(\sigma) \,&\equiv&\,
    \int\limits^{\infty}_{\sigma} d\tau  \, N^{-1}(\tau,0)\,
    Q_{n-\frac{1}{2}}( {\rm ch} \tau)  \label{28}      \\
    G_n(\sigma) \,&\equiv&\,
    \int\limits^{\sigma}_{0} d\tau  \, N^{-1}(\tau,0) \,
    P_{n-\frac{1}{2}}( {\rm ch} \tau) \label{29} \;\;\;,
\eea
and we use the abbreviations $P_{n-\frac{1}{2}} \,\equiv\,
P^0_{n-\frac{1}{2}}$, $\;Q_{n-\frac{1}{2}} \,\equiv\,
Q^0_{n-\frac{1}{2}}$ for the toroidal harmonic functions.

We now turn to the solution of the $\eta$ equation.
Writing (\ref{3_10})  in full
\beq
    \frac{\partial^2 \, \eta}{\partial r^2} \, +\,
    \frac{\partial^2 \, \eta}{\partial z^2} \;=\;
    8\pi G \,k\, \delta(r-a)\,\delta(z) \;\;\;,\label{9955}
\eeq
we can interpret it (after formally replacing $r$ by $x$)
as the potential produced by a uniform thin rod lying in the
$x$-$y$ plane, parallel to the
$y$-axis at $x = a$. This is a standard problem in
electrostatics \cite{P&P},
and in cylindrical coordinates, the solution reads
\beq
    \eta(r,z) \;=\; 2 Gk \log\{[(r-a)^2 \,+\, z^2]/R_0^2\} \;\;\;, \label{9956}
\eeq
where $R_0$ is a constant to be discussed later.

The last remaining metric function to be found is $\zeta$.
Solving ($\ref{3_11}$)  with the above $\eta$, we obtain in
cylindrical coordinates:
\bea
    \zeta(r,z) \,= \,-\eta_0 + \frac{2Gk}{r}\Biggl[
    &r& \log \Bigl(\frac{(r-a)^2+z^2}{R_0^2} \Bigr)
    -a\ \log \Bigl(\frac{(r-a)^2+z^2}{a^2+z^2} \Bigr) \nonumber \\
    &-&2r
    +2 z \Bigl( {\rm tan}^{-1} \bigl(\frac{r-a}{z}\bigr)
              + {\rm tan}^{-1} \bigl(\frac{a}{z} \bigr) \Bigr)
    \Biggr] \;\;\;,
\eea
in which an arbitrary integration function of $z$ has been determined by
the requirement that the solution be regular on the $z$-axis.
Calculating the angular deficit about the $z$-axis yields
\beq
    \delta \phi \, = \, 2 \pi(1-e^{-\eta_0})\;\;\;.
\eeq
Consequently, demanding that there be no conical singularity along the
$z$-axis imposes $\eta_0 = 0$.

Having determined $\nu$, $\eta$, and $\zeta$,
we have formal solutions of the weak-field Einstein equations.
We must now check whether these solutions are valid within the
weak-field approximation. It is clear from (\ref{3_9}), (\ref{3_10})
that $\nu$ and $\eta$ are determined only up to an additive
constant. Moreover, by virtue of (\ref{3_11}), any constant added to
$\zeta$ may be absorbed in a redefinition of $\eta_0$, leaving the
structure of the equation unchanged. This freedom of an additive
constant in each of $\nu$, $\eta$, and $\zeta$ enables one to ensure that
the weak-field approximation is valid at least near the string, the
region of spacetime that will be relevant when investigating conical
singularities and the corresponding deficit angle in the next section.
(In the standard weak-field approximation
in Cartesian coordinates around Minkowski space,
these constants are naturally chosen so that
the potentials vanish at infinity.) Furthermore, all future results
will be seen to be independent of the particular choice of these
additive constants.

Within the model of an infinitely thin string, there is no natural way
to determine the additive constants, since any choice leads to some
of the potentials becoming infinite near the string core, invalidating
the weak-field approximation. (For instance, irrespective of the
value of $R_0$ in (\ref{9956}), $\eta$ is infinite at the ring.)
However, a physical string always has a non-vanishing thickness,
something, strictly speaking, beyond the realm of our thin-string
model. To give, nevertheless, a value to these constants, we may
formally consider a slightly fattened string of core
 radius $R_{\rm core}$,
and fix the additive constants by demanding that the potentials be
small at distances of order $R_{\rm core}$ from the infinitely thin
ring. (For instance, taking $R_0 = R_{\rm core}$ in (\ref{9956})
leads to  $\eta \approx 0$ at distances of the order of $R_{\rm
core}$ from the infinitely thin string.)
In what follows, we stay within the infinitely thin string model,
and we implement the weak-field approximation by the requirement
that $\nu, \eta, \zeta << 1$ on distance scales of order $R_{\rm
core}$ from the infinitely thin string. The actual value selected
for $R_{\rm core}$ does not follow from the thin-string model, and
must be supplied by the underlying physical theory. We emphasize
that we are mathematically dealing with an infinitely thin string
core, with only the faint memory of its physical origin in the
choice of $R_{\rm core}$. It is, however, important to stress that
all future results will be independent of the particular choice of
$R_{\rm core}$.

\setcounter{equation}{0} 

                    \section{Angular Deficit}                                 %

To check the metric (\ref{20}) for conical singularities on the ring
$\sigma = \infty$, we must examine circles $\sigma = \sigma_0$ at
constant $t$ and $\phi$
around the ring, and calculate the ratio of the proper perimeter
to the proper radius in the limit that the proper radius tends to zero.
For  $t$ and $\phi$ constant, the metric (\ref{20}) becomes
\beq
    ds^2\;=\;a^2\,N^{-4}\,e^{2\eta-2\nu}(d\sigma^2\,+ d\psi^2)
    \,\,\,,
    \label{metricc}
\eeq
and the angular deficit $\delta \psi$ is given by
\beq
   \delta\psi \;=\; 2\pi \;-\; \lim_{\sigma_0\rightarrow\infty}\,
   \left[ \int_{-\pi}^\pi d\psi\, (N^{-2}\,e^{\eta-\nu})|_{\sigma
   = \sigma_0} \left/  \int_{\sigma_0}^\infty d\sigma\, N^{-2} \, \right.
   e^{\eta-\nu} \right] \;\;\;. \label{9961}
\eeq
{}From this formula, one sees that the value of $\delta \psi$ is
independent of any additive constants that might appear in $\nu$ or
$\eta$, as emphasised above. In this calculation we are thus free to
ignore such additive constants, and this will be done below.

In order to evaluate the limit appearing in (\ref{9961}), we require
the asymptotic form of $\nu$ and $\eta$ for large values of $\sigma$.
The contribution of $\nu_1$ from (\ref{4_3}) to the
asymptotic behaviour of $\nu$ for $\sigma$ tending to infinity  is readily
found to be
\beq
    \nu_1 \;\rightarrow\; -2G\,(\mu + k)\,\sigma\quad,\quad
    \sigma\rightarrow \infty \;\;\;.\label{9962}
\eeq
(In establishing  this formula, we use the asymptotic
expression $K({\rm th} (\sigma/2))\rightarrow \sigma/2$ when $\sigma$ tends
to infinity \cite{GRAD}.)
In the appendix, we show that   $\nu_2\rightarrow 0$ as
$\sigma \rightarrow \infty$.
Thus, the asymptotic value of $\nu$ does not depend on $\nu_2$, and
this shows  that the radial stresses
${T^r}_r \;=\; (k/r) \, \Theta(r-a)\, \delta(z)$  do not contribute to the
potential near the ring.
Furthermore, the asymptotic behaviour of $\eta$ near the ring is
\beq
    \eta(\sigma,\psi) \rightarrow -4 G k
    \sigma\quad\quad{\rm as} \quad \sigma \rightarrow \infty\;\;\;.
    \label{etanear}
\eeq

With (\ref{9962}) and (\ref{etanear}), it is straightforward to
evaluate the integral
(\ref{9961}) for the deficit angle $\delta \psi$ as
\beq
    \delta \psi \;=\; 4 \pi G \,(\mu - k)\;\;\;, \label{9963}
\eeq
which will be analysed in the conclusion.

The same result for the angular deficit can be obtained in a
different manner, which also sheds light on the spacetime
structure near the ring in the weak-field limit.
Using the asymptotic forms of $\nu$ and $\eta$ near the ring,
the metric (\ref{metricc}) becomes
\beq
    ds^2 \;=\; 4a^2\, e^{-2\sigma} \, e^{+4G(\mu - k)\sigma+2b}
            \,(d\sigma^2\,+\, d\psi^2) \qquad,\qquad  \sigma
            \rightarrow \infty
            \;\;\;,
\eeq
where $b$ denotes the dimensionless combination of all
additive constants appearing in the potentials.
If we now seek coordinates $(\sigma' , \psi')$, so that, near
the ring, the
metric is locally flat,
\beq
    ds^2 \;=\; 4a^2\,e^{-2\sigma'} \,(d\sigma'^{2}\,+\,d\psi'^{2})
    \qquad,\qquad \sigma \rightarrow \infty
     \,\,\,,
    \label{dflat}
\eeq
we see that we must take
\beq
   e^{b}\, e^{-\{(1-\lambda)\sigma\}} d\sigma \;=\;  e^{-\sigma'}
    d \sigma'\,\,\,,
\eeq
where $\lambda \equiv 2G(\mu - k)$.
Thus, to first order in $\lambda$, this yields
\beq
    \sigma' = (1-\lambda)\sigma - \lambda -b\;\;\;.
\eeq
Consequently, if we define $\psi' = (1-\lambda) \psi$, we note that
we do obtain
the metric in the form ($\ref{dflat}$). It is then clear that
$\psi'$ runs from $0$ to $2\pi(1-\lambda)$, and that we thus recover the
deficit angle
\beq
    \delta \psi\,=\,4\pi G(\mu-k)
\eeq
already found in  (\ref{9963}), once again seen to be
independent of the choice of the additive constants.
A similar method was employed by
Vilenkin \cite{3}.

\setcounter{equation}{0} 

                         \section{Conclusion}                                 %

In this paper, we constructed a stationary, axially symmetric
metric satisfying Einstein's  weak-field equations for a source describing
an infinitely thin ring of radius $a$,   linear mass density $\mu$,
and  azimuthal stress per unit length $k$. The
conservation  of stress-energy required the presence of a radial
stress
\beq
    {T^r}_r \;=\;\frac{k}{r}  \, \Theta(r-a) \, \delta(z)
    \label{c1}
\eeq
to support the ring.
The  metric obtained
exhibits a conical singularity  along the ring,
with deficit angle
\beq
    \delta\psi \;=\; 4\pi G (\mu - k) \label{c3}\;\;\;.
\eeq

In the case of a ring of pressureless dust
($k=0$),  a conical singularity  arises, with
deficit angle $4\pi G \mu$. This is usually interpreted by saying
that, because of the presence of the singularity, such a source
cannot remain stationary and must collapse.

When the ring is made of \lq\lq ordinary\rq\rq\ matter
($k > 0\;\;,\;\; k << \mu$),
the conical singularity is still present but is less severe than
for pressureless dust. Such a source
is also considered as
non-physical since, unless the (positive) pressure
equals the energy density, the angular deficit $\delta \psi$
is still non-zero.

When the ring is made of \lq\lq string\rq\rq\ matter ($k=-\mu$),
we recover Vilenkin's result $\delta\psi \; = \; 8 \pi G \mu$.
This result is interesting for two reasons: first, it shows that one
half of the angular deficit of the string source is of \lq\lq
non-string \rq\rq\ origin (since one half of the effect remains for
pressureless dust). Secondly, it
establishes {\it from the field equations}, within the
thin-string model, that
indeed, a conical singularity is
present along a circular string, and that the angular
deficit takes the same value as for a straight string of the
same linear mass density. This was an {\it assumption} made
by FIU  in their investigation of circular
cosmic strings at an instant of time symmetry. Our formalism
provides a weak-field proof of the validity of
the FIU hypothesis.

 \vskip 1 truecm

\section*{Acknowledgements:}
M.V. gratefully acknowledges the
Royal Irish Academy for a grant from the Research Project
Development Fund
and P. Kelleher  for research facilities
provided at the Cork Regional Technical College.
It is a pleasure to  thank  P.A. Hogan, G. Kelly, J.D. Mc Crea, L.
\'O Raifeartaigh  and G. Thomas for enlightening discussions, as well as
the referee for his remarks on an earlier version of related work by
D.Mc M. and M.V.


\setcounter{equation}{0}
\setcounter{section}{0}

\renewcommand{\thesection}{Appendix  A.}
\renewcommand{\theequation}{\Alph{section}.\arabic{equation}}

\section{Asymptotic Behaviour of $\nu_2$ near the Ring}

Establishing the asymptotic behaviour of the potential $\nu_2$
(\ref{27}) near
the ring, namely, for $\sigma$ tending to infinity, is a delicate matter.
This is due to the fact that the source, being
of the form $(k/r)\, \Theta(r-a) \, \delta(z)$, decreases very slowly
at infinity when the radius $r$ increases without limit.  The
difficulty at infinity in $r$ translates into a difficulty at $0$
in $\sigma$.

For the calculation of $F_n(\sigma)$ of (\ref{28}), however,
the point $\sigma = 0$ does not belong to the domain of integration
and one finds, after replacing $Q_{n-\frac{1}{2}}({\rm ch} \sigma)$
by its asymptotic value \cite{QINF} and performing the integration:
\beq
    F_n(\sigma) \;\rightarrow\; (2\pi)^{1/2}\,
    \Gamma(n+\frac{1}{2}) e^{-(n+1)\sigma} / (n+1)! \label{a1}  \;\;\;.
\eeq
On the other hand, a problem does arise in the calculation of
$G_n(\sigma)$: Near $\tau = 0$, the toroidal harmonics
$P_{n-\frac{1}{2}}({\rm ch} \tau)$ tend to $1$ for all $n$ \cite{PZER},
whereas the factor $N^{-1}(\tau,0)$ appearing in the integrand
of (\ref{29}) tends to $2^{1/2}\,\tau^{-1}$.
As a result, $G_n$ diverges, and a cut-off $\varepsilon$ must
be introduced. If this is done, the integrand is finite everywhere,
and the regularised $G_n(\sigma)$, denoted by $G^R_n(\sigma)$,  reads:
\beq
    G^R_n(\sigma) \;\equiv\; \int_\varepsilon^\sigma d\tau
    \, N^{-1}(\tau,0) P_{n-\frac{1}{2}}({\rm ch}\tau) \;\;\;. \label{a2}
\eeq
To go further, one must consider the asymptotic value of
$P_{n-\frac{1}{2}}({\rm ch}\sigma)$.

 For $n = 0$, the corresponding
toroidal harmonic is expressible in terms of the Complete Elliptic
Integral of the First Kind  \cite{PK} as
\beq
    P_{-\frac{1}{2}}({\rm ch}\sigma) \;=\;
    2\pi^{-1} K({\rm th}(\frac{1}{2}\sigma)) /
    {\rm ch}(\frac{1}{2}\sigma)\;\;\;. \label{a3}
\eeq
For $\sigma$ very large, this is asymptotically given \cite{GRAD} by
\beq
    P_{-\frac{1}{2}}({\rm ch}\sigma) \;\rightarrow\;2 \pi^{-1} \, \sigma
    e^{-\sigma/2}\;\;\;. \label{a5}
\eeq
For $n \ge 1$, the asymptotic expression \cite{PINF} of
$P_{n-\frac{1}{2}}({\rm ch}\sigma)$ is
\beq
    P_{n-\frac{1}{2}}\;\rightarrow\;
    \pi^{-1/2} \, (n-1)! \, e^{(n-1/2)\sigma}
    / \Gamma(n+\frac{1}{2})\quad,\quad n\ge 1\;\;\;.\label{a6}
\eeq

When the asymptotic expressions (\ref{a5}), (\ref{a6}) are inserted
into the integral (\ref{a2}) for $G^R_n(\sigma)$, the following
results are obtained:
\renewcommand{\arraystretch}{1.7}
\bea
    G^R_0(\sigma) \;&\rightarrow&\; A(\varepsilon) \quad,\quad
    A(\varepsilon) \equiv \int_\varepsilon^\infty
    d\tau N^{-1}(\tau, 0) P_{-\frac{1}{2}}({\rm ch}\tau) \label{a8} \\
    G^R_1(\sigma) \;&\rightarrow&\; (\frac{2}{\pi})^{1/2} \;
    \sigma \;/\Gamma(\frac{3}{2}) \label{a10} \\
    G^R_n(\sigma) \;&\rightarrow&\; (\frac{2}{\pi})^{1/2}\; (n-2)! \;
    e^{(n-1)\sigma} / \Gamma(n+\frac{1}{2}) \quad,\quad n \ge 2
    \;\;\;, \label{a12}
\eea
so that, finally, the asymptotic form of the series (\ref{27})
becomes
\beq
    \nu_2 \;\rightarrow\; -Gk \left[ 2^{1/2} \pi A(\varepsilon) \;+\;
    o(e^{-\sigma})\right]\;\;\;.\label{a14}
\eeq

As one can see, for every non-vanishing value of $\varepsilon$,
the potential $\nu_2$ tends to a constant on the ring, that is, for
$\sigma$ tending to infinity. The fact that $\nu_2$ is
asymptotically constant is sufficient to establish the result
(\ref{9963}) of the deficit angle, irrespectively
of the value of the constant. It is, however, possible
to go one step further and renormalise the potential so that
$\nu_2$ vanishes at the ring. This amounts to subtracting from
$\nu_2$ the $(\sigma,\psi)$-independent term
 $2^{1/2} \pi  A(\varepsilon)$. (Although, when the cut-off
$\varepsilon$ tends to $0$, the constant $A(\varepsilon)$ diverges,
this procedure is nothing more than a conventional
renormalisation.)

\rule{13.3cm}{0.5pt}

\begin{tabbing}
       ${}^{\ast}$\= \kill
       ${}^{\ast} $On leave from Lyman Laboratory of Physics,
                   Harvard University,\\
       \>Cambridge, MA 02138, U.S.A. \\   
       ${}^{\dag}$Address after September 1, 1992: Dept. of
       Mathematics, Statistics, and Computer \\
        \>Science, Dalhousie University,
       Halifax, Nova Scotia, Canada B3H 4H6. \\
       ${}^{\ddag}$Research Associate of the Dublin Institute for
                 Advanced Studies.
\end{tabbing}


\end{document}